\documentstyle[prl,aps,epsf]{revtex}
\begin{document}
\draft
\twocolumn[
\hsize\textwidth\columnwidth\hsize\csname@twocolumnfalse\endcsname
\preprint{}
\title{
Quasiparticle States around a Nonmagnetic Impurity in D-Density-Wave 
State of High-$T_c$ Cuprates
}
\author{Jian-Xin Zhu,$^{1}$ Wonkee Kim,$^{2}$ C.S. Ting,$^{1}$ 
and J. P. Carbotte$^{2}$} 
\address{$^{1}$ Texas Center for Superconductivity and Department of 
Physics, University of Houston, Houston, TX 77204\\
$^{2}$Department of Physics \& Astronomy, Mcmaster University, Hamilton, 
Ontario, Canada, L8S 4M1
}
\maketitle
\begin{abstract}
{ 
Recently Chakravarty {\em et al.} proposed an ordered $d$-density wave 
(DDW) state as an explanation of the pseudogap phase in underdoped 
high-temperature cuprates. We study the competition between the 
DDW and superconducting ordering based on an effective mean-field 
Hamiltonian. We are mainly concerned with the effect of the DDW 
ordering on the electronic state around a single nonmagnetic impurity. 
We find that a single subgap resonance peak appears in the local 
density of state around the impurity. In the unitary limit, the 
position of this resonance peak is always located at $E_r=-\mu$ with 
respect to the Fermi energy. This result is dramatically different from  
the case of the pure superconducting state 
for which the impurity resonant energy is 
approximately pinned at the Fermi level. This can be used to probe the 
existence of the DDW ordering in cuprates.    
}
\end{abstract} 
]

\narrowtext
Recently a $d$-density wave (DDW) state was proposed to model
the pseudogap in the underdoped high-$T_c$ cuprates ~\cite{Chakr01a,Nayak00}.
The key feature of the DDW state is
staggered orbital magnetic moments (i.e., staggered currents) 
which break parity and 
time-reversal symmetry as well as the translation
invariance by one lattice constant and 
rotation by $\pi/2$.  This proposal of a new order 
parameter differs fundamentally from other theoretical 
ideas~\cite{Wen96,Ivanov00,Leung00} for which the staggered current is a 
fluctuating rather than a static quantity. 
Whether such a DDW state really exists in the 
phase diagram of high-$T_c$ cuprates is an interesting theoretical
question~\cite{Wang00a}. More recently, the 
SU(2)~\cite{Lee00} or U(1)~\cite{Han00}
mean-field theory of the $t$-$J$ model 
predicted that the DDW order could exist in a vortex core.  

An interesting question is the effect of a single 
nonmagnetic impurity on the DDW state. The single impurity problem 
in the superconducting state has been intensively studied both 
theoretically~\cite{Bala95,Salk96,Byers93,Zhu00a} and 
experimentally~\cite{Pan00,Huds99,Yazd99}. However, the same problem has 
not been well studied in the pseudogap (PG) state, which has been  
indicated by 
STM~\cite{Renner98} and by ARPES experiments~\cite{Loes96}. 
Currently, the origin of the PG state remains unclear. Experimentally it 
was also argued~\cite{Kras00} that in the underdoped cuprates the 
superconducting and pseudo gap coexist unlike
in the pre-formed pair model~\cite{Emery95}.
The electronic states around a single impurity may serve as a 
local probe to distinguish between these scenarios. 
Within the $T$-matrix approximation, it was argued~\cite{Krui00} that, as 
long as the band density of states is depleted at the Fermi energy, the 
existence of  
impurity resonant states is robust regardless of the microscopic 
origin of the PG state. Note that, without invoking any particular model, 
the band density of states was assumed in Ref.~\cite{Krui00} to be 
symmetric with respect to the 
Fermi energy. This choice corresponds to the chemical potential  
$\mu$ being zero in a two-dimensional tight-binding model with 
nearest-neighbor hopping. In addition, because of the lack of a 
specific model, the calculation of the local density of states (LDOS) 
around an impurity goes beyond the scope of that work. 
The purpose of this paper is to address this
question within the DDW state, for which 
the LDOS around the impurity is calculated. 
Our main result is that, in the strong scattering limit, 
the position of the induced 
resonant state has a one-to-one correspondence to the chemical potential, which 
is dramatically different from the case of the superconducting state (and also 
of the pre-formed pair scenario), where the resonant energy is 
almost 
pinned at the Fermi energy. This prediction can be readily used by the STM 
experiments to detect for the DDW state in the underdoped cuprates.  
 
We start with a phenomenological model
including the intersite charge density wave order in a $d$-wave
superconductor. One could also 
envision intersite spin density wave instead of
the charge density order; however, as long as the density order is to be
interpreted as a pseudogap the charge density order is more
reasonable experimentally~\cite{Chakr01a}.
The effective
Hamiltonian is written as:
\begin{eqnarray}
H&=&\sum_{{\bf ij},\sigma} [-t+(-1)^{\bf i}W_{\bf ij}]
c_{{\bf i}\sigma}^{\dagger}c_{{\bf j}\sigma}
+\sum_{{\bf i},\sigma}(U_{\bf i}-\mu)c_{{\bf i}\sigma}^{\dagger}
c_{{\bf i}\sigma}
\nonumber \\ 
&&+\sum_{\bf ij} (\Delta_{\bf ij} c_{{\bf i}\uparrow}^{\dagger}
c_{{\bf j}\downarrow}^{\dagger} +\Delta_{\bf ij}^{*} c_{{\bf j}\downarrow}
c_{{\bf i}\uparrow})\;.
\label{EQ:MFA}
\end{eqnarray}
Here $c_{{\bf i}\sigma}^{\dagger}$ creates an electron with spin $\sigma$
at site ${\bf i}$. The summation is over the nearest neighbors sites.
$t$ is the hopping integral and $\mu$ is the chemical potential.
$U_{\bf i}$ is introduced to model the potential scattering, if any, from
impurities or defects.
$W_{\bf ij}$ and $\Delta_{ij}$ 
are, respectively, the
DDW and DSC order parameters, which are
determined by the self-consistent conditions:
\begin{equation}
W_{\bf ij}=(-1)^{\bf i}\frac{V_{DDW}}{2}\langle c_{{\bf i}\sigma}^{\dagger}
c_{{\bf j}\sigma} -c_{{\bf j}\sigma}^{\dagger}c_{{\bf i}\sigma}\rangle\;,
\label{ansatz}
\end{equation}
and
\begin{equation}
\Delta_{\bf ij}=\frac{V_{DSC}}{2}\langle
c_{{\bf i}\uparrow}c_{{\bf j}\downarrow}
-c_{{\bf i}\downarrow}c_{{\bf j}\uparrow}\rangle\;,
\end{equation}
where $V_{DDW}$ and $V_{DSC}$ are the interaction for the DDW and DSC
channels. 
$(-1)^{\bf i}$ in the ansatz Eq.~(\ref{ansatz}) ensures that nesting takes
place at half filling, and that $W_{\bf ij}$ is pure-imaginary.
A. Ghosal {\it et. al.} \cite{ghosal}
have considered a Fock shift, which is similar
with $W_{\bf ij}$ but does not have the site-dependent
factor $(-1)^{\bf i}$.
A possible model for the Hamiltonian \ref{EQ:MFA} is
the $t$-$J$ model. Applying the mean-field approximation
and the ansatz for the density wave to the exchange interaction,
one can obtain density and
superconducting order. Introduction of an intersite interaction
in addition to the exchange interaction gives different values of
$V_{DSC}$ and $V_{DDW}$.

The mean-field Hamiltonian (~\ref{EQ:MFA}) can be diagonalized by solving 
the
resulting Bogoliubov-de Gennes equations self-consistently
\begin{equation}
\sum_{\bf j} \left(
\begin{array}{cc}
{\cal H}_{\bf ij}& \Delta_{\bf ij} \\
\Delta_{\bf ij}^{*} & -{\cal H}_{\bf ij}^{*}
\end{array}
\right)
\left(
\begin{array}{c}
u_{\bf j}^{n} \\
v_{\bf j}^{n}
\end{array}
\right)
=E_{n}
\left(
\begin{array}{c}
u_{\bf i}^{n}   \\
v_{\bf i}^{n}
\end{array}
\right)\;,
\label{EQ:BdG}
\end{equation}
where the single particle Hamiltonian ${\cal H}_{\bf ij}= 
-t\delta_{{\bf i}+\boldmath{\mbox{$\delta$}},{\bf j}}
+(-1)^{\bf i}W_{\bf ij} +(U_{\bf i}-\mu)\delta_{\bf ij}$. 
and the self-consistency now reads
\begin{equation}
W_{\bf ij}=(-1)^{\bf i} \frac{iV_{DDW}}{2} \sum_{n}
\mbox{Im}[u_{\bf i}^{n}u_{\bf j}^{n*}+v_{\bf i}^{n}v_{\bf j}^{n*}]
\tanh \left( \frac{E_{n}}{2k_{B}T} \right)\;,
\end{equation}
and
\begin{equation}
\Delta_{\bf ij}=\frac{V_{DSC}}{2}\sum_{n}
(u_{\bf i}^{n}v_{\bf j}^{n*} +v_{\bf i}^{*}u_{\bf j}^{n})\tanh
\left( \frac{E_{n}}{2k_{B}T}\right) \;.
\end{equation}

{\em Competition between DDW and DSC orderings.} 
The phase diagram with the DDW and DSC orderings in Ref.~\cite{Chakr01a} 
was constructed  
from experiments. Therefore, it is important to self-consistently
study the phase diagram of a clean system with DDW and DSC 
channel interactions.
Although the DDW state breaks the
translational invariance by one lattice constant $a$, it is invariant 
under the translation with spacing $\sqrt{2}a$ along the diagonal 
directions
of the square lattice. Hereafter we will measure the length in units of
$a$ and the energy in unit of $t$. 
The site-dependent factor $(-1)^{\bf i}$
enables us to consider the square lattice
as a bipartite lattice.
Choosing the basis cell with the size
$\sqrt{2} \times \sqrt{2}$, we solve the BdG equation (~\ref{EQ:BdG}) and
find the eigenvalues $E_{\bf k}=\pm E_{1,2}({\bf k})$
with $E_{1,2}({\bf k})=[(\xi_{\bf k} \mp \mu)^{2} +\Delta_{\bf 
k}^{2}]^{1/2}$, where $\epsilon_{\bf k}=-2(\cos 
k_{x}+\cos k_y)$, $W_{\bf k}=2iW_d(\cos k_x -\cos k_y)$, 
$\xi_{\bf k}=\sqrt{\epsilon_{\bf k}^{2}+\vert W_{\bf k}\vert^{2}}$
 and $\Delta_{\bf k}=2\Delta_d
(\cos k_x -\cos k_y)$.
The eigenfunctions corresponding to $E_{1,2}({\bf k})$
are, $(u_{\bf k}e^{i\phi_{DDW}},\pm u_{\bf k},v_{\bf 
k}e^{-i\phi_{DSC}},\pm v_{\bf 
k}e^{-i(\phi_{DSC}-\phi_{DDW})})^{Transpose}$. 
Here $u_{\bf k}^{2}=\frac{1}{4}\left( 1 \pm \frac{\xi_{\bf k} \mp 
\mu}{E_{1,2}({\bf k})}\right)$ and 
$v_{\bf k}^{2}=\frac{1}{4}\left( 1 \mp \frac{\xi_{\bf k} \mp 
\mu}{E_{1,2}({\bf k})}\right)$, 
and $\phi_{DDW}=\tan^{-1}(-iW_{\bf k}/\epsilon_{\bf k})$ 
and  $\phi_{DSC}$ are the phase associated with the DDW and DSC 
orders. Substitution of these eigenfunctions into the self-consistent 
equations for DDW and DSC order parameter leads to:
\begin{eqnarray}
W_d&=&\frac{V_{DDW}}{4N_{L}}\sum_{{\bf k},\alpha}
 \frac{W_{d}(\cos k_x-\cos k_y)^{2}}{\xi_{\bf k}}
\frac{\xi_{\bf k}+(-1)^{\alpha}\mu }{E_{\alpha}}
\nonumber \\
&&\times \tanh \frac{E_{\alpha}}{2k_{B}T}\;,
\end{eqnarray}
\begin{equation}
\Delta_{d}=\frac{V_{DSC}}{4N_{L}}
\sum_{{\bf k},\alpha} 
\frac{\Delta_{d}(\cos k_x-\cos k_y)^{2}}{E_{\alpha}}
\tanh \frac{E_{\alpha}}{2k_{B}T}\;,
\end{equation}
where $\alpha=1,2$ and $N_{L}$ is the number of the basis cells.
Notice that at the fixed doping, the chemical potential itself is 
determined from $\delta=1-n$ where the filling factor
$n=\frac{2}{N_{L}}\sum_{{\bf k},\alpha} \{ 
f(E_{\alpha})u_{\bf k}^{2} +[1-f(E_{\alpha})]v_{\bf k}^{2} \}$.
The solutions to the above equations have shown that the critical temperature 
$T_{DDW}^{0}$ for the ``bare'' transition to the DDW state decreased more 
quickly than $T_{DSC}^{0}$ for the DSC state. Therefore, only 
when the DDW channel interaction is larger than the DSC one, can there 
exist separated regions where either the DDW or DSC state is dominant. 
As a model calculation, we display in Fig.~\ref{FIG:PHASE} the 
temperature versus doping phase diagram for $V_{DDW}=1.6$ and 
$V_{DSC}=1.4$.  In the region of doping close to zero, 
the DDW state is  dominant while in the region of $\delta$
away from zero, the 
DSC state is dominant.  In addition, we have a finite area of 
coexisting DDW+DSC phase. 
The size of the pure DDW region could be increased with a different
choice of $V_{DDW}$ and $V_{DSC}$ but this is not expected to lead
to important quantitative change. Also half filling in our model is
not an insulating anti-feromagnetic phase because we have not included
any of this physics in our model so that the pseudogap state extends
down to $\delta=0$.

This competition can be understood from a 
phenomenological Ginzburg-Landau (GL) theory. The GL free energy density, 
in terms of both the DDW and DSC order parameters, can be constructed from
a symmetry analysis:
\begin{eqnarray}
f&=&\alpha_{DSC} \vert 
\Delta_{d} \vert^{2} + \alpha_{DDW } \vert W_d \vert^{2} 
+\beta_{1}\vert \Delta_{d}\vert^{4}+\beta_{2}\vert W_d \vert^{2} 
\nonumber \\
&&+\beta_{3} \vert \Delta_d \vert^{2} \vert W_d \vert^{2} 
+\beta_{4} (\Delta_d^{*2} W_{d}^{2} +\Delta_{d}^{2} W_{d}^{*2} )\;,
\end{eqnarray} 
where we assume $\alpha_{DSC}=\alpha_{DSC}^{\prime}(T-T_{DSC}^{0})$ and 
$\alpha_{DDW}=\alpha_{DDW}^{\prime}(T-T_{DDW}^{0})$, and $\beta$'s are 
all positive. As an example, in the region where 
$T_{DDW}^{0}>T_{DSC}^{0}$, the second phase transition temperature for the 
DSC ordering is renormalized by the pre-existing DDW order parameter: 
\begin{equation}
T_{DSC}=T_{DSC}^{0}-
\frac{(\beta_3-2\beta_4)(T_{DDW}^{0}-T_{DSC}^{0})}
{2\beta_2\alpha_{DSC}^{\prime}/\alpha_{DDW}^{\prime}-(\beta_3-2\beta_4)}\;.
\end{equation}
Therefore, even if the ``bare'' critical temperatures for 
both ordering are very close to each other, the second-phase transition 
temperature for the appearance of DSC (or DDW) order parameter can be strongly 
suppressed by the dominant DDW (or DSC) order parameter.
At the mean-field level, we conclude that when $V_{DDW}>V_{DSC}$, the 
phase diagram is consistent with that proposed in Ref.~\cite{Chakr01a}.
Here we also would like to point out that the U(1)
mean-field theory of the $t$-$J$ model gives
rise to $V_{DDW}=0.5J$ while $V_{DSC}=J$, which may explain the absence of
the bulk DDW state in that model~\cite{Wang00a}.

{\em The LDOS around a single nonmagnetic impurity.}
The LDOS is given by:
\begin{equation}
\rho_{\bf i}(E) = -2\sum_{n}[\vert 
u_{\bf i}^{n}\vert^{2} 
f^{\prime}(E_{n}-E) +\vert v_{\bf 
i}^{n}\vert^{2}f^{\prime}(E_{n}+E)]\;, 
\end{equation}
where a factor $2$ arises from the summation over spin, and $f^{\prime}(E) 
\equiv df(E)/dE$ with the Fermi distribution 
function $f(E)=1/(e^{E/k_BT}+1)$. 
The LDOS $\rho_{\bf i}(E)$ is proportional to 
the local differential tunneling conductance which can be 
measured by STM experiments~\cite{Tinkham75}.
In our numerical calculation, we take the supercell size 
$N_x=N_y=32$ and the number of supercells $M=6 \times 6$.
Since we are most interested in identifying the qualitative difference 
in the electronic states around the impurity for the DDW  
and DSC states, the thermally broadening effect will not 
be considered here and the temperature is fixed at $T=0.01$. 
For simplicity, we also use uniform order parameter as 
an input to diagonalize Eq.~(\ref{EQ:BdG}) and  the suppression of 
the order parameter near the impurity is ignored. 
This approximation should be acceptable to answer the questions 
we ask here.  

We first consider the case of weak or moderately 
strong impurity scattering. Figure~\ref{FIG:LDOS-0} gives a plot of
the LDOS spectrum
directly on the single-site impurity (a) and on a site one lattice constant 
away (b) for a pure DDW state. When the impurity scattering is weak 
($U_0=4$), the LDOS on both the  impurity site and on its nearest neighbor 
has a single peak below the Fermi energy. 
Although the LDOS at the impurity site is similar to the 
case of a weak nonmagnetic impurity in a pure DSC 
state~\cite{Salk96,Zhu00b}, the LDOS displays a single resonant peak 
at the nearest neighbor site while the DSC shows
a double-peak structure with one peak above 
and the other below the Fermi energy. When the impurity scattering 
becomes stronger ($U_0=100$), the subgap resonant peak in the 
LDOS at the impurity site is shifted toward $E_r=0$ 
with respect to the Fermi energy at the same time its amplitude at the 
impurity site is  strongly suppressed because of the 
strong impurity scattering while
at the nearest neighbor site it is strongly enhanced. 

In the unitary limit, the energy position $E_r$ for the impurity resonant 
state in a pure DDW state is very sensitive to the chemical potential. 
In Fig.~\ref{FIG:LDOS-1}, the LDOS spectrum is plotted at the 
nearest-neighbor site for various values of 
$\mu$. We see in Fig.~\ref{FIG:LDOS-1} that the resonant energy 
position is exactly equal to $E_r=\vert \mu\vert$ ($\mu<0$), which is in 
sharply contrast to the case of a single impurity in a pure DSC state, where 
the resonant energy is not sensitive to the chemical potential
and roughly zero (i.e., very close to the Fermi surface). 
This difference can be understood as follows: 
The energy dispersion for the pure DDW state  
$E_{1,2}({\bf k})=\vert \sqrt{\epsilon_{\bf k}^{2}+\vert W_{\bf k}\vert^{2}}
\mp \mu \vert$ shows that the overall quasiparticle 
band is shifted by $\mu$ because of nature 
of the DDW state. Correspondingly, the resonant peak which always 
exists at the band center is shifted by $\mu$.
However, due to the 
pairing mechanism, the quasiparticle for the pure 
DSC state, as given by $E_{1,2}({\bf k})=[(\vert \epsilon_{\bf k}\vert \mp 
\mu)^{2} +\Delta_{\bf k}^{2}]^{1/2}$,  
are always excitated w.r.t. the Fermi surface instead of the band 
center. Consequently, the impurity induced resonant state has the energy 
almost pinned around the Fermi surface.
Moreover, as in the case of a DSC 
state, we have found 
that the impurity resonant peak intensity of the LDOS in a pure DDW state 
exhibits the Friedel-like spatial oscillation: It has local maxima on the 
sublattice containing  the nearest neighbors and local 
minima on the sublattice containing the impurity site itself. 

Finally, in Fig.~\ref{FIG:LDOS-2}, the LDOS spectra 
at the nearest-neighbor site of a moderately strong impurity is plotted     
in a mixed state of DDW and DSC ordering. For a comparison, the LDOS 
spectrum at the same site is also displayed in a pure DSC state. 
In the pure DSC state, we have the double-peak structure with the intensity of 
the peak above the Fermi energy stronger than that below the Fermi 
energy. However, in the mixed state with both orderings present, 
the intensity of 
the peak below the Fermi energy is stronger than that above 
because the impurity scattering in the DDW order shifts the resonant 
peak below the Fermi energy. 

In conclusion, we have studied the competition between the DDW and DSC 
ordering. The implication of the DDW state for the impurity resonant 
state has been discussed in detail. The qualitative difference 
found in the 
resonant state in the DDW ordering compared with 
that in the DSC state can be used 
as a smoking gun for the existence of the DDW state in high-$T_c$ 
cuprates. 
Experimentally, the existence of this state can be identified by (i)
detecting whether the local differential tunneling conductance 
near a single weak or moderately strong nonmagnetic impurity 
exhibits a single rather than double subgap structure around the Fermi 
surface and (ii) detecting the position of the subgap resonant peak in the 
conductance near a unitary nonmagnetic impurity which will have a strong doping 
dependence. An STM measurement is most suitable for this test.

{\bf Acknowledgments}:
J.Z wish to thank A.V. Balatsky and Q.H. Wang
for useful discussions, and W.K aknowledges A. Ghosal for helpful
discussions.
This work was
supported by the Texas Center  for Superconductivity at the University of
Houston through the State of
Texas (J.Z and C.S.T), by the National Science and Engineering Research
Council of Canada (W.K and J.P.C).

\begin{figure}
\caption[*]{The temperature versus doping$(\delta)$
phase diagram obtained from 
the Hamiltonian~(\ref{EQ:MFA}) with channel interaction $V_{DDW}=1.6$ 
and $V_{DSC}=1.4$.
}
\label{FIG:PHASE}
\end{figure}

\begin{figure}
\caption[*]{The LDOS spectrum at the impurity site (a) and at its nearest 
neighbor (b) in a pure DDW state for various strength of the impurity 
scattering $U_0=4$ (red-solid line) and  $U_0=100$ (green-dashed line). 
Also shown the LDOS spectrum (blue-dash-dotted line) at a position far 
away from the impurity.
The other parameter values: $W_{0}=0.1$, $\mu=0$, and  $T=0.01$.
The blue line in the bulk density of states for reference.
}
\label{FIG:LDOS-0}
\end{figure}

\begin{figure}
\caption[*]{The LDOS spectrum at the  
nearest neighbor of a unitary impurity 
in a pure DDW state for various values of the chemical potential 
$\mu=0$ (red-solid line), $-0.1$ (green-dashed line), and $-0.2$ 
(blue-dash-dotted line). The other parameter values: $W_{d}=0.1$, 
$U_0=100$, and  $T=0.01$.
}
\label{FIG:LDOS-1}
\end{figure}

\begin{figure}
\caption[*]{The LDOS spectrum at the  nearest neighbor of a weak 
impurity in a mixed state of DDW and DSC orderings (red-solid line).
The other parameter values: $W_{d}=0.08$, $\Delta_d=0.1$, $U_0=4$, 
$\mu=0$, and  $T=0.01$. Also shown is the spectrum in a pure DSC state 
with $\Delta_d=0.1$ (green-dashed line). 
}
\label{FIG:LDOS-2}
\end{figure}


\begin{references}
\bibitem{Chakr01a} S. Chakravarty {\em et al.}, Phys. Rev. B {\bf 63}, 
094503 (2001).

\bibitem{Nayak00} C. Nayak, Phys. Rev. B {\bf 62}, 4880 (2000); {\em 
ibid.} {\bf 62}, R6135 (2000).

\bibitem{Wen96} Xiao-Gang Wen and P. A. Lee, Phys. Rev. Lett. {\bf 76}, 
503 
(1996).

\bibitem{Ivanov00} D. A. Ivanov, P. A. Lee, and Xiao-Gang Wen, Phys. Rev. 
Lett. {\bf 84}, 3958 (2000).

\bibitem{Leung00} P. W. Leung, cond-mat/0008419.



\bibitem{Wang00a} Qiang-Hua Wang, Jung Hoon Han, and Dung-Hai Lee, 
cond-mat/0011398.

\bibitem{Lee00} P. A. Lee and Xiao-Gang Wen, cond-mat/0008419;
Jun-ichiro Kishine, P. A. Lee, and Xiao-Gang Wen, cond-mat/0103148.


\bibitem{Han00} Jung Hoon Han, Qiang-Hua Wang, and Dung-Hai Lee, 
cond-mat/0012450; Qiang-Hua Wang, Jung Hoon Han, and Dung-Hai Lee,
cond-mat/0102048.

 

\bibitem{Bala95} A. V. Balatsky, M. I. Salkola, and A. Rosengren, Phys.
Rev. B {\bf 51}, 15547 (1995).

\bibitem{Salk96} M. I. Salkola, A. V. Balatsky, and D. J. Scalapino,
Phys. Rev. Lett. {\bf 77}, 1841 (1996).

\bibitem{Byers93} J. Byers, M. E. Flatt\'{e}, and D. J. Scalapino, Phys. 
Rev. Lett. {\bf 71}, 3363 (1993); M. E. Flatt\'{e}, Phys. Rev. B {\bf 61}, 
R14920 (2000).

\bibitem{Zhu00a} Jian-Xin Zhu and C. S. Ting, cond-mat/00012276, and 
references therein.

\bibitem{Pan00} S. H. Pan {\em et al.}, Nature (London) {\bf 403}, 746
(2000).

\bibitem{Huds99} E. W. Hudson {\em et al.}, Science {\bf 285}, 88 (1999).

\bibitem{Yazd99} A. Yazdani {\em et al.}, Phys. Rev. Lett. {\bf 83}, 176
(1999).

\bibitem{Renner98} Ch. Renner {\em et al.}, Phys. Rev. Lett. {\bf 80}, 149 
(1998).

\bibitem{Loes96} A. G. Loeser {\em et al.}, Science {\bf 273}, 325 (1996); 
M. R. Norman {\em et al.}, Nature {\bf 392}, 157 (1998).

\bibitem{Kras00} V. M. Krasnov {\em et al.}, Phys. Rev. Lett. {\bf 84}, 
5860 (2000).

\bibitem{Emery95} V. J. Emery and S. A. Kivelson, Nature {\bf 374}, 434 
(1995).

\bibitem{Krui00} H.V. Kruis, I. Martin, and A.V. Balatsky, 
cond-mat/0008349. 

\bibitem{ghosal} A. Ghosal {\it et. al.}, Phys. Rev. B {\bf 63},
020505 (2000).

\bibitem{Tinkham75} M. Tinkham, {\em Introduction to
Superconductivity} (McGraw Hill, New York, 1975).

\bibitem{Zhu00b} Jian-Xin Zhu, C. S. Ting, and Chia-Ren Hu, Phys. Rev. B 
{\bf 62}, 6027 (2000).

\end{references}
\end{document}